# Ponderomotive tunneling in electron-laser interactions

Ju Gao* and Fang Shen
*University of Illinois, Urbana, IL 61801*
(Dated: February 11, 2007)

It is known that the ponderomotive potential from the laser electron interaction behaves like a static potential that scatters off electrons. We have investigated the process of the electron tunneling the ponderomotive potential by using the nonperturbative quantum electrodynamic approach with a modified Gaussian quantized laser field. The results show that the ponderomotive tunneling is significantly larger than the tunneling of a static potential of the same profile. The tunneled electron is discrete in energy spectrum and spatial diffraction as the result of multiphoton processes. A resonant tunneling occurs when the ponderomotive energy equals the multiple integer of the laser photon energy.

*Introduction* — Ponderomotive energy ($U_p$) emerges as an important concept in many strong laser field effects [1]. It originates from the non-oscillating part of the quadratic term of the vector potential $\mathbf{A}(\mathbf{r},t) = \mathbf{A}\cos(\mathbf{k}\cdot\mathbf{r} - \omega t)$ and is expressed by

$$U_p(eV) \equiv \frac{e^2 \mathbf{A}^2}{4mc^2}. \qquad (1)$$

The classical theory interprets the ponderomotive energy as an averaged energy of the wiggling electron inside the laser field [2] [3] [4]; the quantum interpretation is that it is the total energy of the photons that dress the electron [5]. It is evident from above threshold ionization (ATI) and high order harmonic generation (HHG) experiments that the electron experiences an effective static-like potential. When the spatial aspect of the laser intensity profile is considered, the ponderomotive potential gives rise to a force according to $d\mathbf{P}/dt = -\nabla U_p$. The effect of such force is first discovered by Bucksbaum *et al* [6] in the ATI experiment where the ponderomotive potential alters the electron angular distributions. Later they report a direct study of the ponderomotive potential effect [7] where a free electron, generated by ATI, intercepts the ponderomotive potential transversely as illustrated by Fig. 1. The experiment shows that the electron can be accelerated or decelerated by the ponderomotive potential much the same way as a static potential, which can be utilized for laser manipulation of the matters [8].

So far the discussions about ponderomotive scattering in the literature have assumed a Newtonian electron. If the wave nature of the electron is considered, ponderomotive tunneling, a process that may already take place in experiments, should accompany ponderomotive scattering. In a layman's term, if the ponderomotive potential is used as a barrier to trap electrons, then the barrier may be leaky due to the tunneling effect.

Most tunneling effects, however, take place in dimensions much smaller than the typical extend of the pon-deromotive potential [9]. Should the ponderomotive potential behave exactly the same as a static potential, considering the typical laser beam size, it will be too wide for the electron to tunnel through with a meaningful probability. However, we can not conclude the ponderomotive tunneling is negligible without considering the role of photons: the ponderomotive potential is made up of photons that can be absorbed by electrons. The more energetic electron can clear the ponderomotive potential with near unity probability. In that case, the total tunneling probability reflects mostly the multi-photon absorption probability. The challenge now is to find a theoretical framework that will produce numerical results to be compared with the experiments.

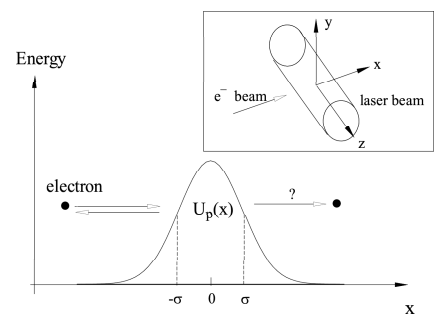

FIG. 1: An electron intercepting a laser in transverse direction observes ponderomotive scattering and tunneling effects. The inset illustrates the coordinate where the origin is at the center of the laser focus. For numerical discussion, the laser wavelength is chosen to be 1.064$\mu$m and width $\sigma = 6\mu$m, and the electron energy $E_0 = .54$ eV.

The uniqueness of the ponderomotive potential and the involvement of multiphoton process prevent us from using the existing methods [9] to treat the ponderomotive tunneling problem. Our approach is based on the nonperturbative quantum electrodynamics (NPQED) developed in recent years for the high field interaction [10–12]. However, plane wave is always assumed for the laser field in NPQED that is not suitable for the discussion of pon-deromotive tunneling. For the first time in literature, a Gaussian mode quantized field Volkov solution is derived that depicts the electron and laser interaction. Formulae that describe the ponderomotive tunneling are then derived based on the formal scattering theory. Numerical calculation is carried out to show salient characteristics of the ponderomotive tunneling.

---
*Electronic address: jugao@uiuc.edu

*Theoretical Development* The scattering state for the electron and laser interaction process is described by the Lippmann-Schwinger equation,

$$\Psi_i^+ = \phi_i + \frac{1}{E_i - H_0 + i\epsilon} V \Psi_i^+, \quad (2)$$

where $\phi_i$ is the photon electron state before the interaction and is also the eigenstate of the noninteracting Hamiltonian $H_0$ with eigenenergy $E_i$. $V$ is the interaction Hamiltonian. The energy uncertainty $\epsilon$ is related to the transition time of the interaction $T$ by $\epsilon = \frac{\hbar}{T}$.

The probability of finding the electron in a free state $\phi_f$ after the interaction is determined by projecting the scattering state $\Psi_i^+$ onto $\phi_f$, $\Omega_{fi} \equiv \langle \phi_f | \Psi_i^+ \rangle$, where $\Omega_{fi}$ is the Moller operator.

The differential transition rate of the electron tunneling the laser field is expressed by:

$$\frac{dW}{V_e \left(\frac{1}{2\pi\hbar}\right)^3 d^3 \boldsymbol{P}_f} = \frac{4}{T} |\Omega_{fi}|^2, \quad (3)$$

where $\mathbf{P}_f$ is the final electron momentum and $V_e$ is the normalization volume of the electron.

Now inserting a set of quantized field Volkov solution $\Phi_\mu$ to the equation $(H_0 + V)\Phi_\mu = E_\mu \Phi_\mu$, we have

$$\Omega_{fi} = \Sigma_\mu \langle \phi_f | \Phi_\mu \rangle \langle \Phi_\mu | \phi_i \rangle \frac{\epsilon^2 + i\epsilon(E_\mu - E_i)}{(E_\mu - E_i)^2 + \epsilon^2}. \quad (4)$$

In the previous NPQED treatment, only the Volkov state on the energy shell $E_\mu = E_i$ is included. Here we sum over Volkov states of all energy levels, weighted by the profile function $\frac{\epsilon^2 + i\epsilon(E_\mu - E_i)}{(E_\mu - E_i)^2 + \epsilon^2}$, because of the energy uncertainty caused by the finite laser beam size. The initial and final electron photon states are free of the interaction and expressed by

$$\phi_i = \frac{1}{\sqrt{V_e}} e^{iP_{xi}x/\hbar} e^{iP_{yi}y/\hbar} e^{iP_{zi}z/\hbar} | l \rangle;$$
$$\phi_f = \frac{1}{\sqrt{V_e}} e^{iP_{xf}x/\hbar} e^{iP_{yf}y/\hbar} e^{iP_{zf}z/\hbar} | m \rangle \quad (5)$$

where $P_{xi} = \sqrt{2mE_0}, P_{yi} = 0, P_{zi} = 0$ are the initial electron momenta and $P_{xf}, P_{yf}, P_{zf}$ are the final electron momenta. $|l\rangle$ and $|m\rangle$ are the initial and final photon states with $l$ and $m$ number of photons, respectively.

The foremost task is to find the solution of $\Phi_\mu$ that satisfies

$$\{\hbar\omega N_a + \frac{\hbar^2}{2m}[-i\nabla - \frac{e}{c}\boldsymbol{A}(x,y,z)]^2\}\Phi_\mu = E_\mu \Phi_\mu, \quad (6)$$

where $\mathbf{A}(x,y,z) = gf(x,y)\mathbf{x}[ae^{-ikz} + a^\dagger e^{ikz}]$ is the vector potential of the quantized field. $N_a$, $a$ and $a^\dagger$ are photon number, annihilation and creation operators, respectively. $g = \sqrt{\frac{\hbar}{2\epsilon_0 V_\gamma \omega}}$ is the normalization factor of the laser field where $V_\gamma$ is the normalization volume of the field. $f(x,y) = e^{-4\ln 2 \frac{x^2+y^2}{\sigma^2}}$ describes the Gaussian field distribution where we neglect the dependence on $z$ coordinate. We choose the laser to be linearly polarized along $\mathbf{x}$ without loosing generality because the ponderomotive potential is independent of the laser polarization. Throughout the paper all formulae are expressed in SI unit, instead of atomic or natural units that are commonly adopted in quantum mechanical theories, for the convenience of numerical calculation.

Exact solution of the Gaussian quantized field Volkov state (Eq. 6) can not be found. A key step to solve the problem is understanding that the Gaussian function $f(x,y)$ varies much slower than the wavefunction $\Phi(x,y)$ itself. Physically this requires that the dimension of the laser beam is much larger than the de Broglie wavelength of the electron, which is true throughout this investigation. Under this approximation, operators $\frac{\partial}{\partial x}$ and $\frac{\partial^2}{\partial x^2}$ commute with the function $f(x,y)$ but not with the wavefunction $\Phi(x,y)$. Thus the solution of $\Phi_\mu$ is found to be,

$$\Phi_\mu = \frac{1}{\sqrt{L_z}} e^{-ikzN_a} e^{ip_z z/\hbar} e^{-F_2} e^{-F_1} \varphi(x,y) |n\rangle, \quad (7)$$

where $L_z$ is the normalization length in z direction, $p_z$ is the total momentum of the system and $|n\rangle$ is the photon state with $n$ photons. We will present the detailed derivation elsewhere but give the expression of the transformation,

$$\begin{aligned} e^{-F_1} &= e^{\delta^* a - \delta a^\dagger}; \\ e^{-F_2} &= e^{\chi a^2/2 - \chi a^{\dagger 2}/2}, \end{aligned} \quad (8)$$

where

$$\begin{aligned} \chi &= \frac{1}{2} \text{Tanh}^{-1} \frac{-\frac{\hbar^2}{m} \frac{e^2 g^2}{c^2} f^2(x,y)}{[\hbar\omega + \frac{\hbar^2}{m} \frac{e^2 g^2}{c^2} f^2(x,y)]}; \\ \delta &= -\frac{\hbar e g}{mc} f(x,y)(\cosh\chi + \sinh\chi) \frac{1}{2C}(-i\hbar\frac{\partial}{\partial x}); \\ C &= \frac{1}{2}\hbar\omega\cosh 2\chi + \frac{\hbar^2}{2m} \frac{e^2 g^2}{c^2} f^2(x,y)(\sinh 2\chi + \cosh 2\chi). \end{aligned} \quad (9)$$

Therefore the transformation $e^{-F_2} e^{-F_1}$ contains both the electronic and photonic operators.

The electronic wavefunction in Eq. 7 satisfies

$$\{\frac{P_z^2}{2m} + \frac{\hbar^2}{2m}[-\frac{\partial^2}{\partial x^2} - \frac{\partial^2}{\partial y^2}] + n\hbar\omega + U_p(x,y)\}\varphi(x,y)|n\rangle$$
$$= E_\mu \varphi(x,y)|n\rangle, \quad (10)$$

where $U_p(x,y) \equiv \frac{\hbar^2}{2m} \frac{e^2 g^2}{c^2} f^2(x,y)n = f^2(x,y)U_p$. Assuming the electron beam is narrower than the laser beam and aims at the center of the laser, we may separate the y variable $\varphi(x,y) = X(x)\frac{1}{\sqrt{L_y}}exp(iP_y y/\hbar)$ where $L_y$ is the normalization length in y direction. $X(x)$ is then found by solving Eq. 10 with WKB method,

$$X(x) = \frac{1}{D}\sqrt{\frac{1}{P_x(x)}}e^{i\int_{-\infty}^x P_x(x')dx'/\hbar}, \quad (11)$$

where

$$P_x(x) = \sqrt{2m\{E_\mu - [\frac{P_z^2}{2m} + \frac{P_y^2}{2m} + n\hbar\omega + U_p(x)]\}}, \quad (12)$$

is the x-axis momentum inside the laser field.

The quantized Gaussian field Volkov state is finally obtained after evaluating the operators:

$$\Phi_\mu = \frac{1}{\sqrt{L_z}}\frac{1}{\sqrt{L_y}}e^{iP_y y/\hbar}e^{-ikzN_a}e^{i(P_z+u_p\hbar k+n\hbar k)z/\hbar}$$
$$\sum_{j_1,j_2=-\infty}^{\infty}(-i)^{j_1}J_{j_1}[\eta(x)]J_{j_2}[-\frac{u_p(x)}{2}]X(x)|n+2j_2+j_1> \quad (13)$$

where $P_z$ is the z-axis electron momentum during the interaction and $\eta(x) = \sqrt{2}\frac{\sqrt{2mU_p(x)}}{m\hbar\omega}P_x(x)$, $u_p(x) = U_p(x)/(\hbar\omega)$ are the coordinate-dependent variables of the Bessel functions.

Now the Moller operator $\Omega_{fi}$ (Eq. 4) can be evaluated by using Eqs. 13 and 5, which reveals the momenta along z direction

$$\begin{aligned}P_{zf} &= j''\hbar k;\\ P_z &= j\hbar k - u_p\hbar k,\end{aligned} \quad (14)$$

where $j = l - n$ is the number of photons absorbed by the electron entering the field and $j'' = l - m$ is the net photon number absorbed during the whole process. $j''$ is not necessary the same as $j$ because the electron can re-emit some photons back to the laser field when leaving the field.

The energy conservation must be satisfied before and after the interaction:

$$E_i = l\hbar\omega + \frac{P_{xi}^2}{2m} = E_f = m\hbar\omega + \frac{P_{xf}^2}{2m} + \frac{P_{zf}^2}{2m}, \quad (15)$$

which by combining with Eq. 15 gives the final electron momentum in x direction:

$$P_{xf} = \sqrt{2m[\frac{P_{xi}^2}{2m} + j''\hbar\omega - \frac{P_{zf}^2}{2m}]}. \quad (16)$$

For a fixed net absorption of $j''$ photons, the transition rate is expressed by,

$$\begin{aligned}W &= \frac{4}{T}|\Omega_{fi}|^2\\ &= \frac{4}{T}\left|\Sigma_{j,E_\mu}m_{fi}(j,j'';E_\mu)\frac{\epsilon^2 + i\epsilon(E_u-E_i)}{(E_u-E_i)^2+\epsilon^2}\right|^2,\end{aligned} \quad (17)$$

where the matrix element

$$m_{fi}(j,j'';E_\mu) = \frac{1}{L_x}$$
$$\sum_{\substack{j_2=-\infty;\\j_1=j-2j_2}}^{\infty}(-i)^{j_1}\left\langle e^{-iP_{xi}x/\hbar}J_{j_1}[\eta(x)]J_{j_2}[-\frac{u_p(x)}{2}]X(x)\right\rangle \sum_{\substack{j_2'=-\infty;\\j_1'=j'-2j_2'}}^{\infty}(-i)^{j_1'}\left\langle e^{iP_{xf}x/\hbar}J_{j_1'}[\eta(x)]J_{j_2'}[-\frac{u_p(x)}{2}]X^*(x)\right\rangle, \quad (18)$$

where $j' = j - j''$ and $L_x$ is the normalization length in the x direction.

*Numerical examples.* The result of the theoretical development described above is used in evaluating the example as illustrated by Fig. 1. The transition rate (Eq. 17) is calculated by first calculating the probability $|\Omega_{fi}|^2$ and then divided by the transition time $T = 2\sigma/v_0 = 6.9 \times 10^{-11}$ s for $v_0 = 4.4 \times 10^5$m/s at $E_0 = 0.54$ eV.

Figure 2 shows the tunneled electron energy spectra at different laser intensities. The electron tunnels through the potential with appreciable transition rate at discrete energies $E_0 + j''\hbar\omega$ that reaches the maximum at the energy just above the ponderomotive potential, and then decays rapidly at higher photon order $j''$, resembling other multiphoton processes. The point to make here is that the tunneling is the result of the electron picking up photons from the same laser that forms the pondero-motive potential. The contribution from the tunneling terms without the assist of the photon is negligible. It is to be noted that $j''$ represents the net photon absorbed by the electron and the actual photons involved are much more because the electron can absorb multiple photons and then re-emit during the process.

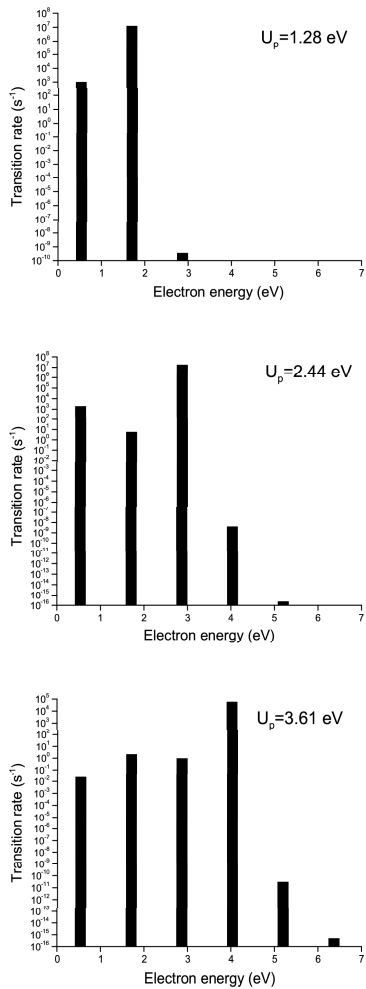

FIG. 2: The ponderomotive tunneling electron displays discrete energy spectra at different $U_p$. The parameters are the same as in Fig. 1. Notice the laser photon energy $\hbar\omega = 1.16$ eV

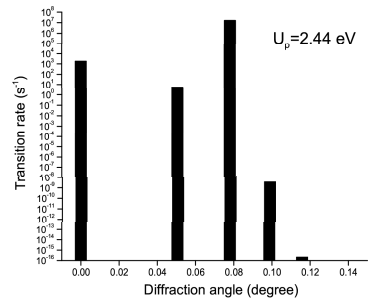

FIG. 3: The ponderomotive tunneling electron is diffracted from the single laser field. Unlike KD effect, the diffraction is not symmetrical, appearing only on the positive side of z-axis.

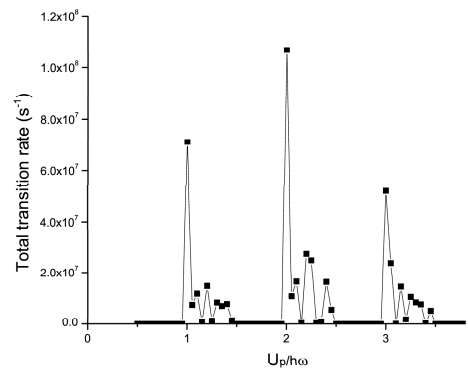

FIG. 4: The total electron transition rate is plotted against the normalized ponderomotive energy $U_p/\hbar\omega$.

The expression of the final electron momentum suggests that the tunneled electron is spatially diffracted along the z direction. The effect should be linked to Kapitza-Dirac (KD) effect that was predicted [13] and observed [14] for an electron diffracting off a standing wave laser field. Figure 3 shows that the diffraction can take place for electrons with one laser field.

In an earlier paper, we have discussed a resonant intensity effect [12] involving ATI electrons leaving the laser field. The resonance comes when the corresponding ponderomotive energy is equal to the integer multiples of photon energy due to the balance between the photons and ponderomotive four momentum [4]. Such resonant intensity effect should also be expected for the process discussed here that includes the electron entering and leaving the field. Figure 4 plots the total electron transition rate for all tunneled electrons as a function of the ponderomotive energy and the resonant effect is shown. In Ref [12], we assume the laser field is a plane wave that results in divergent resonance. Since the Gaussian laser field is used here, the divergence is eliminated and the resonant peaks are broader.

*Conclusion* We have used the nonperturbative QED to study electron tunneling of ponderomotive potentials. The effect extends the research on basic electron-laser scattering, e.g. Compton scattering or Kapitza-Dirac diffraction, where the electron energy is always assumed to be higher than the ponderomotive energy. The key conclusion is that the electron can tunnel the ponderomotive potential by absorbing the photons from the laser field. The tunneled electron is spatially diffracted and discrete in energy spectrum. A resonant tunneling occurs when the ponderomotive energy approaches the multiple integer of the laser photon energy.